\documentclass[twocolumn,showpacs,amsmath,amssymb]{revtex4} 

\makeatletter

\usepackage{graphicx}
\usepackage{dcolumn}
\usepackage{bm}

\begin{document}

\title{Hofstadter Butterfly Diagram in Noncommutative Space}

\author{Hidenori Takahashi} 
\email{htaka@phys.ge.cst.nihon-u.ac.jp}
\author{Masanori Yamanaka}
\email{yamanaka@phys.cst.nihon-u.ac.jp}

\affiliation{Department of Physics, College of Science and Technology,
Nihon University, Kanda-Surugadai 1-8-14, Chiyoda-ku,
Tokyo 101-8308, Japan}

\date{\today}

\begin{abstract}
We study an energy spectrum of electron moving under the constant
magnetic field in  two dimensional noncommutative space. It take place with the gauge 
invariant way. 
The Hofstadter butterfly diagram of the noncommutative space 
is calculated in terms of the lattice model which
 is derived by the Bopp's shift for space and by the Peierls
substitution for external magnetic field.

We also find the fractal structure in new diagram.
Although the global features of the new diagram are similar to the
diagram of the commutative space, the detail structure is different from it.
\end{abstract}


\maketitle

The recent development of the noncommutaitve space (NC) physics 
has shown the interesting results in the theoretical point of view.
Most development has been made in the string theory
\cite{conn_doug_schw,sei_wit,douglas_rev}.
The string theory is believed to be a ruling theory 
of the quantum gravity.
In the quantum gravity, space and time coordinates should be
fluctuating.
As the result there exist some scale $\theta$ in the theory. 
Therefore, space and time coordinate should be discretize in units 
of $\theta$.
This means that the space of the quantum gravity should be described 
in terms of new geometry instead of the Riemann geometry.

The noncommutative geometry is the leading candidate for the geometry 
of quantum gravity and is naturally formulated in terms of D-branes.
The coordinates of this geometry satisfy the commutation relation,
\begin{equation}
  [\hat x_\mu, \, \hat x_\nu ]_\star = i\theta_{\mu \nu}, 
  \quad \theta_{\mu \nu} = - \theta_{\nu \mu}
\end{equation}
where $\theta_{\mu \nu}$ is a noncommutative parameter 
which is related to the string tension $\alpha'$.
In noncommutative space, the product between functions becomes 
$\star$-product;
\begin{align}
 f(x) \star g(x) 
  &= \exp \left[\dfrac{i}{2} \theta_{\mu \nu} 
                \dfrac{\partial}{\partial \xi^\mu}
                \dfrac{\partial}{\partial \zeta^\nu}
         \right]
     f \left(x+\xi \right) g \left(x+\zeta \right)
                \biggl|_{\xi=\zeta=0} \nonumber \\
  &= f(x)g(x) + \dfrac{i}{2} \theta_{\mu \nu} \partial_\mu 
                       f \partial_\nu g + \textrm{O}(\theta^{2})
\end{align}
For example, two dimensional noncommutative coordinate 
$\hat{\bm{r}} =(\hat x, \hat y)$ satisfies
\[
  \left[\hat x, \, \hat y \right]_\star 
   = \hat x \star \hat y - \hat y \star \hat x 
   = i \theta,
\]
where we take $\theta_{\mu \nu} = \theta \epsilon_{\mu \nu}$
and $\epsilon_{\mu \nu}$ is two rank antisymmetric tensor with
$\epsilon_{12}=1$. Further,
\[
  \left[\hat x, \, \hat x \right]_\star = 0,
  \quad \left[\hat y, \, \hat y \right]_\star = 0.
\]
This means that a particle has nontrivial phase shift 
only when the electron comes back to starting point
with different path in the noncommutative space.

The noncommutative space ($\hat{\bm{r}}$) is reformulated 
in terms of commutative coordinate ($\bm{r}$) with the Bopp's shifts
\cite{curt_fair_zach}. 
\begin{equation}
 \hat x = x - \dfrac{\theta}{2} p_y, 
 \quad \hat y = y + \dfrac{\theta}{2} p_x ,
\label{noncomm_comm_rel}
\end{equation}
where $x, y, p_x, p_y$ satisfy the following commutation relations,
\[ [x, \, p_x]= i, \quad  [y, \, p_y]= i,  \quad [x, \, y]= 0, 
\quad  [p_x, \, p_y]= 0 . \]
In the literature, the noncommutative parameter can 
be interpreted to an external magnetic field.
This is an intersting analogy between the noncommutative parameter
and the external magnetic field.
But our approach is distinct from it.
In this Letter, we consider electron in the noncommutative space with 
external noncommutative magnetic
field.

In the noncommutative space, U(1) gauge theory (QED) becomes a 
noncommutative U(1) gauge theory (NCQED).
The NCQED model has a local  NC U(1) gauge symmetry. The NC gauge 
transformation of NC fermion $\hat \psi$
and the NC gauge field $\hat A_\mu$ is given by
\begin{equation}
\hat \psi' = e^{-i\lambda}_\star \star \hat \psi,
\end{equation}
where
\begin{equation}
U = e^{-i \lambda }_\star \equiv 1-i\lambda +
\dfrac{(-i)^2}{2!}\lambda 
\star \lambda  +  \cdots
\end{equation}
and
\begin{equation}
\hat A_\mu' = U \star A_\mu \star U^\dagger + \dfrac{i}{e} U \star 
\partial_\mu U^\dagger.
\end{equation}
It is important to note that the NC magnetic field is not a gauge
invariant object like as QCD because
the (NC) field strength $\hat F_{\mu \nu}$ is not gauge invariant.
Therefore, we must take care of the gauge invariance of the physical 
observable more carefully in NCQED than in QED.

It is interesting to enlarge the application of the noncommutative
theory to quantum mechanics
\cite{curt_fair_zach,duval_horv,nair_pol,fal_gam_loe_roj_prd66,chain_pres_shek_ture,dayi_jellal}, 
e.g. Landau level problem 
\cite{duval_horv,nair_pol,dayi_jellal,fal_gam_loe_roj_prd66}. 
The continuous spectrum of an free electron becomes discrete spectrum
when the constant magnetic field is switched on. Further, to take a
lattice 
formulation,
we find the fractal structure in the Hofstadter butterfly diagram 
\cite{hofstadter,hiramoto_kohmoto_rev}.

In the noncommutative space, the Hamiltonian of the electron should be 
given by
\begin{equation}
H_{nc} = \dfrac{1}{2m} \left(  \hat{\bm{p}} + e \hat{\bm{A}} 
(\hat{\bm{r}})\right)^2.
\end{equation}
In terms of the Bopp's shift formulation, the NC U(1) gauge field 
$\hat{\bm{A}}$ is mapped
to the gauge field $\bm{A}$ in the commutative space 
\cite{fal_gam_loe_roj_prd66} as
\begin{equation}
\hat{A}_j \rightarrow A_j - \dfrac{1}{2}\theta_{a b} \partial_a A_j
p_b .
\end{equation}
Therefore, the Hamiltonian can be rewritten in terms 
of the commutative coordinates as
\begin{equation}
H_{nc} =\dfrac{1}{2m} \left(p_j + e A_j 
        - \dfrac{e}{2}\theta_{a b} \partial_a A_j p_b\right)^2
\end{equation}
To take the symmetric gauge,
\begin{equation}
 \bm{A} = \left( -\dfrac{B}{2}y, \, \dfrac{B}{2} x, 0 \right),
\label{symmetric_gauge}
\end{equation}
the Hamiltonian becomes \cite{dayi_jellal}
\begin{equation*}
 H = \dfrac{1}{2m} \left(\Pi_x^2 + \Pi_y^2 \right),
\end{equation*}
where
\begin{align*}
 \Pi_x = \xi_\theta^{-2} p_x - \dfrac{eB}{2} y, \quad
 \Pi_y = \xi_\theta^{-2} p_y + \dfrac{eB}{2} x 
\end{align*}
and
\begin{equation}
\xi_\theta^{-2} \equiv 1-\dfrac{eB\theta}{4}.
\end{equation}
In this case, the commutation relations become
\begin{subequations}
\begin{equation}
\left[x, \, \Pi_x \right] = i\xi_\theta , 
\quad  \left[y, \, \Pi_y \right] = i\xi_\theta, 
\end{equation}
\begin{equation}
\left[\Pi_x, \, \Pi_y \right] = -ieB \xi_\theta^{-2} .
\label{noncmm_pp_rel}
\end{equation}
\end{subequations}
Further, the Hamiltonian can be expressed in terms 
of a harmonic oscillator \cite{dayi_jellal},
\begin{equation}
H = \dfrac{eB}{m} \xi_\theta^{-2}
    \left(a^\dagger a + \dfrac{1}{2} \right),
    \quad \left[a, \, a^\dagger \right] = 1
\end{equation}
with the transformation,
\begin{equation*}
\Pi_x = \dfrac{\xi_\theta^{-1}}{i}\sqrt{\dfrac{eB}{2}}
   \left(a-a^\dagger \right), \quad
 \Pi_y = \xi_\theta^{-1}\sqrt{\dfrac{eB}{2}}
   \left(a+a^\dagger \right) .
\end{equation*}
This means that the energy spectrum of the electron in the
noncommutative space is discretized 
and the Landau level like spectrum is also emerged in noncommutative
space.
The gap of the energy spectrum is given by 
\begin{equation}
 E = \dfrac{eB}{m} \xi_\theta^{-2} .
\label{noncmm_energy}
\end{equation}
Note that the Landau level in the noncommutative space 
is different from the commutative case by a factor $\xi_\theta^{-2}$.
That is, it is expected that the difference of the Hofstadter diagram 
between the commutative and the noncommutative case emerge.

In the above, however, we chose the symmetric gauge to derive the 
Hamiltonian.
In the gauge theory, a physical observable must be gauge invariance 
quantity.
In the commutative space, we can easily verify 
that the spectrum of the Landau level and the relation
\begin{equation}
 [\Pi_x, \Pi_y] \psi = -i eB \psi
\label{comm_pp_rel}
\end{equation}
are gauge invariance. On the other hand, it is not obvious that the 
commutation relation eq.(\ref{noncmm_pp_rel})
and the energy eq.(\ref{noncmm_energy}) are gauge invariant.

It can be shown in the order of $\theta (\theta \ll 1)$ that after the 
gauge transformation
\begin{equation}
D_\mu' \star \psi' = e^{-i \lambda} \star \bigl(D_\mu \star \psi
\bigr),
\end{equation}
where $D_\mu =\partial_\mu + ieA_\mu$ is the covariant derivative.
Therefore, we can find that
\begin{equation}
D_\mu' \star \bigl(D_\nu' \star \psi' \bigr) 
= e^{-i \lambda} \star 
\biggl(D_\mu \star \bigl(D_\nu \star \psi \bigr) \biggr) .
\label{noncomm_DD_gi}
\end{equation}
It is important to note here that the order of the star product 
with a differential operator $\partial_\mu$ is important.
From eq.(\ref{noncomm_DD_gi}), we can easily prove that
\begin{equation}
[D_\mu, D_\nu]_\star \diamond \psi = c_0 \psi
\end{equation}
is NC U(1) gauge invariant when $c_0$ is constant.
Here, for convenience, we introduced $\diamond$ operation
between $\Phi_1 (\equiv \phi_1 \star \phi_2)$ and $\Psi$ as
\begin{equation}
\Phi \diamond \Psi 
\equiv \biggl(\phi_1 \star \bigl(\phi_2 \star \Psi \bigr)\biggr) .
\end{equation}
To take the symmetric gauge eq(\ref{symmetric_gauge}), we have the
relation
\begin{equation}
[\Pi_x, \Pi_y]_\star \diamond \psi 
  = - ieB \left(1- \dfrac{eB \theta}{4} \right) \psi
\label{nccomm_dd_sg}
\end{equation}
where $\bm{\Pi} \equiv - i \bm{D}$. Further, the NC Schr\"odinger
equation of the electron whose
energy is $E_0$,
\begin{equation}
 H \diamond \psi = E_0 \psi, \qquad 
H =-\dfrac{1}{2m} \bm{D} \star \bm{D}
\end{equation}
is gauge invariant. Accordingly, the commutation relation
eq.(\ref{nccomm_dd_sg}) and the energy of electron  is invariant under
the NC U(1) gauge transformation.
More detail discussion of the gauge invariance will be found in 
\cite{taka_yama}

Finally, we should conclude that the electron moving under the
constant 
magnetic field $B$ 
in the noncommutative space is equivalent to the electron moving
under the constant magnetic field $B \xi_\theta^{-2}$ in the
commutative space,
\begin{equation}
 eB \longrightarrow eB \left(1- \dfrac{eB \theta}{4} \right).
\label{nccomm_mag}
\end{equation}

Now, we consider the Hofstadter butterfly diagram in the
noncommutative space.
The Hofstadter butterfly diagram is the relation between the energy 
spectrum of lattice
Hamiltonian of electron and the flux $\phi=eB$.
It gives the Landau level in the continuous limit
\cite{thouless_83prb}.

Electrons in a two-dimensional lattice 
subject to a uniform magnetic field show extremely rich 
and interesting behavior.
Usually a solid with well localized atomic orbits
is modeled by the tight-binding Hamiltonian 
and the effect of the magnetic field is included
by the Peierls substitution.
To characterize the system, one can use the flux per plaquette,
$\phi$. 
When $\phi$ is rational, the single-particle Schr\"odinger equation
is reduced to the Harper's equation \cite{harper,wannier}.
The equation appears in many different physical contexts 
ranging from the quasiperiodic systems \cite{hiramoto_kohmoto_rev} 
to quantum Hall effects \cite{tknn,ntw}.
When $\phi$ is irrational, the spectrum is known to have
an rich structure like the Cantor set and to exhibit
a multifractal behavior \cite{hofstadter,hiramoto_kohmoto_rev}.

The lattice version of the Hamiltonian of the electron can be
formulated from the continuum Hamiltonian in terms of the Bopp's shift
for 
space and
the Peierls substitution for the external magnetic field.
In this respect, the lattice Hamiltonian can be described by
\begin{align}
 H = T_x + T_y + T_x^\dagger + T_y^\dagger, 
\end{align}
where $T_\mu$ is translation operator which is defined as
\[
T_\mu = e^{i \sqrt{\frac{a}{4m}} \Pi_\mu},
\]
where $a$ is lattice spacing. We can verify that
\begin{equation}
 T_x(a) T_y(b) = e^{iab/(\ell^2 \xi_\theta^2)} T_y(b) T_x(a) ,
\end{equation}
where $\ell^{-2}=eB$. Therefore, the path ordering integral of the 
plaquette
\[ \textrm{P} \exp \left(i \oint_C \!dx_\mu \,\Pi_\mu \right) 
  = T_x(a) T_y(b) T_x^{-1}(a) T_y^{-1}(b) \]
becomes
\begin{equation}
 \textrm{P} \exp \left(i \oint_C \!dx_\mu \,\Pi_\mu \right) 
   = \exp \left(  i\dfrac{2 \pi\Phi}{\varphi_0 \xi_\theta^2} \right) ,
\end{equation}
where $\varphi_0 \equiv 2 \pi/e$ is a flux quantum.
This means that the number of the magnetic flux 
which is penetrated to each plaquette is
\begin{equation}
 N = \dfrac{\Phi}{\varphi_0 \xi_\theta^2} .
\end{equation}

Further, to take the tight-binding approximation and the Peierls 
substitution, we can write the translation operator as \cite{hatsugai}
\begin{align*}
 T_x &= \sum_{m,n} c^\dagger_{m+1,n} c_{m,n} e^{i \Theta_{m,n}^x},\\
 T_y &= \sum_{m,n} c^\dagger_{m,n+1} c_{m,n} e^{i \Theta_{m,n}^y}
\end{align*}
with
\begin{equation}
 \textrm{rot}_{(m,n)} \Theta = \Delta_x \Theta_{m,n}^y - \Delta_y 
\Theta_{m,n}^x = 2 \pi (\ell \xi_\theta)^{-2},
\label{phase}
\end{equation}
Finally, we obtain the tight-binding model 
\begin{eqnarray}
 H= & &   \sum_{m,n} c^\dagger_{m+1,n} c_{m,n} e^{i \Theta_{m,n}^x} 
\nonumber\\
    & & + \sum_{m,n} c^\dagger_{m,n+1} c_{m,n} e^{i \Theta_{m,n}^y} +
    h.c.
\end{eqnarray}
with (\ref{phase}). For convenience, we take the Landau gauge
\begin{eqnarray}
\Theta_{m,n}^x=0 , \ \ \ 
\Theta_{m,n}^y=2\pi\phi\left(1-\dfrac{\phi\theta}{4}\right)m,
\end{eqnarray}
where $\phi=eB$. 

In noncommutative space, the system is characterized by two
parameters, the scale parameter $\theta$ and the flux $\phi$.
We numerically calculate the energy spectrum as a function 
of the flux $\phi$ for a fixed $\theta$
to see how the Hofstadter butterfly diagram is modified
in the presence of the parameter $\theta$.

\begin{figure}[htb]
\scalebox{0.4}{
\rotatebox{0}{\includegraphics{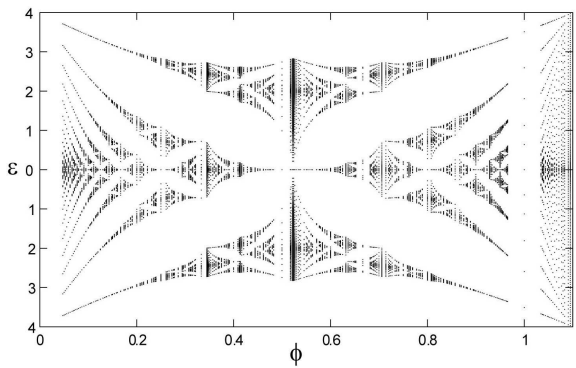}}
}
\caption{Noncommutative Hofstadter butterfly diagram 
at $\theta=1/3$ which corresponds to the one period
of commutative diagram.}
\label{Hofstadter_short}
\end{figure}

\begin{figure}[htb]
\scalebox{0.4}{
\rotatebox{0}{\includegraphics{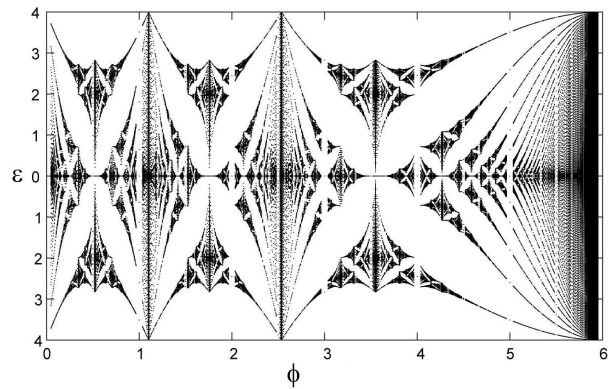}}
}
\caption{Noncommutative Hofstadter butterfly diagram at $\theta=1/3$.}
\label{Hofstadter}
\end{figure}

The energy dispersion relation $\epsilon(k_x, k_y)$ 
for a rational flux, $\phi=p/q$ 
where $p$ and $q$ are mutually prime number,
is given by the equation
\begin{eqnarray}
\det\left(
\begin{array}{ccccc}
M_1       & e^{ik_x}  &           &           & e^{-ik_x} \\
e^{-ik_x} & M_2       & e^{ik_x}  &   0       &           \\
          & \ddots    & \ddots    & \ddots    &           \\
          &   0       & \ddots    & M_{q'-1}   & e^{ik_x}  \\
e^{ik_x}  &           &           & e^{-ik_x} & M_{q'}       \\
\end{array}
\right) =0. 
\label{seculareq}
\end{eqnarray}
where
\[ M_n=2\cos(k_y +\phi') -\epsilon(k_x, k_y)
\]
and
\[
\phi' \equiv \phi\left(1- \dfrac{\phi\theta}{4}\right)
= \dfrac{p'}{q'}.  \]
We numerically diagonalize (\ref{seculareq})
and find a Hofstadter butterfly diagram of the noncommutative space.
We show the results in Fig.\ref{Hofstadter_short} and
Fig.\ref{Hofstadter} for $\theta=1/3$ and $\phi=p/q$ with 
$p=1,2,\cdots,6q-1$, $2 \le q <29$,
and a restriction that $q'<10000$. 

Fig.\ref{Hofstadter_short} shows the region 
of $0< \phi <6-2 \sqrt{6}$.
The new diagram is different from the diagram 
of the commutative space.
There is no periodicity with respect to $\phi$.
Indeed, the detail structure is different from it.
Fig.\ref{Hofstadter} shows the wide area of $\phi$.
From Fig.\ref{Hofstadter}, we can find that new diagram 
is stretched quadratically to the $\phi$-axis direction 
and the difference between them becomes sharp.

On the other hand, the global energy distribution in the new diagram
has a self-similar (fractal) structure.
In this sense, there is similar structure 
between the commutative and the noncommutative diagram.

In the commutative case, there are $q$ sub-bands
at a rational flux, $p/q$.
For small $q$, the band width of each sub-bands are large.
In the non-commutative space, most of the same region
corresponds to that with an irrational flux.
Then the energy spectrum becomes singular,
i.e. the number of the sub-bands diverges 
and the band width is zero.
The rational flux in commutative space
and the flux $\phi$ in noncommutative space relates 
\begin{eqnarray}
\frac{p}{q}=\phi\left(1-\dfrac{\frac{r}{s}\phi}{4}\right).
\end{eqnarray}
for a rational $\theta=r/s$,
where $r$ and $s$ are mutually prime number.
It is easy to show that the quadratic equation 
\begin{eqnarray}
qr\phi^2-4sq\phi+4sp=0
\end{eqnarray}
hardly have rational solution.

For example, we have two wide sub-bands at $1/2$ flux
in commutative space.
The corresponding value in non-commutative space
is $\phi=6-\sqrt{30}$.
It is a irrational limit of the external magnetic field.
Therefore, no sub-band with a wide band width exists.

\end{document}